\documentclass[final]{svjour2}
\usepackage{graphicx}
\usepackage{rotating}
\usepackage{amssymb}
\usepackage{mathptmx}
\usepackage[numbers]{natbib}
\usepackage{mathtools}

\makeatletter
\journalname{Journal of Low Temperature Physics}

\bibpunct{}{}{,}{s}{}{,}

\begin{document}

\newcommand{\hdblarrow}{H\makebox[0.9ex][l]{$\downdownarrows$}-}
\title{Vortex reconnections in anisotropic trapped
three-dimensional Bose-Einstein condensates}

\author{T. Wells$^{1,2}$ $^\star$ \and A. U. J. Lode$^3$ \and V.S. Bagnato$^2$\and M.C. Tsatsos$^2$}

\institute{1:Department of Applied Mathematics and Theoretical Physics, University of Cambridge, Cambridge CB3 0WA, United Kingdom.
\\2: Instituto de F\' isica de S\~ao Carlos, Universidade de S\~ao Paulo, Caixa Postal 369, 13560-970 S\~ao Carlos, S\~ao Paulo, Brazil.
\\3: Department of Physics, University of Basel, Klingelbergstrasse 82, CH-4056 Basel, Switzerland
\\$\star$ \email{tjw78@cam.ac.uk}}

\date{\today}

\maketitle

\keywords{ultracold bosons, many-body, MCTDHB, http://ultracold.org.}

\begin{abstract}

Quantum vortex reconnections can be considered as a fundamental unit of interaction in complex turbulent quantum gases. Understanding the dynamics of single vortex reconnections as elementary events is an essential precursor to the explanation of the emergent properties of turbulent quantum gases. It is thought that a lone pair of quantum vortex lines will inevitably interact given a sufficiently long time. This paper investigates aspects of reconnections of quantum vortex pairs imprinted in a Bose-Einstein condensate held in an anisotropic three dimensional trap using an exact many-body treatment. In particular the impact of the interaction strength and the trap anisotropy in reconnection time is studied. It is found that interaction strength has no effect on reconnection time over short time scales and that the trap anisotropy can cause the edge of the condensate to interfere with the reconnection process. It is also found that the initially coherent system fragments very slowly even for relatively large interaction strength and therefore the system likes to stay condensed during the reconnections.
\end{abstract}


\section{Introduction}
Bose-Einstein condensation was first predicted theoretically in 1924 by S.N. Bose and A. Einstein for noninteracting bosonic particles. Its first experimental realization was only achieved in 1995 in interacting dilute atomic vapours, by E. Cornell and C. Wieman\cite{BEC2} and independently by R. Hulet\cite{BEC3} and W. Ketterle\cite{BEC4}. This peculiar finding has since sparked significant research interest in ultracold bosonic gases and their dynamics. Among the numerous unique features of BECs are the superfluid properties. Specifically, the nucleation and interaction of quantized vortices and turbulent behavior are of exceptional interest. Quantized vorticity follows from the single-valuedness of the superfluid wavefunction and plays a crucial role in the dynamics of the BECs\cite{Feynman1955}. The understanding of quantum turbulence, i.e. the chaotic interaction of numerous quantum vortex lines, has long been sought and remains an active area of research \cite{Vinen2002,Henn2009}. 

Quantum vortices were first observed experimentally in BECs in 1999-2000\cite{firstVorts,secondVorts}. Decades earlier, the Gross-Pitaevskii equation (GPE) had been devised in order to describe vortices in superfluids\cite{Gross1961,fetter}. A vortex reconnection is a  general phenomenon in which two vortex lines meet at a point and exchange tails. In 1987 Ashurst and Meiron simulated the reconnections with a Biot-Savart model\cite{A&M} and later on, in 1993, Koplik and Levine showed that quantum vortex reconnections (QVRs) appear in the dynamics of the GPE\cite{K&L}. In 1994 Waele and Aarts with a Biot-Savart model claimed a universal route to reconnection for all kinds of initial vortex-antivortex arrangements\cite{W&A}.  More recent relevant work includes the study of vortex reconnections in untrapped superfluids\cite{Tebbs2010}, in trapped BECs at finite temperatures\cite{joy}, the computation of the minimum distance between two approaching vortices\cite{zuccher} and the calculation of the energy spectra of gases with reconnecting vortices\cite{Nemirovskii}.

Contemporary approaches to quantum turbulence focus on drawing classical analogies and contrasting them with the quantum case. This can be considered a top-down approach to understanding turbulent dynamics, using what is already known about classical dynamics to understand quantum dynamics. Today the GPE is the foremost basis for studies of quantum turbulence. Yet, as a mean-field theory it neglects crucial quantum effects such as fragmentation and system correlations or even the angular symmetries of the isotropic gas\cite{Tsatsos2010}. An alternative approach to understanding quantum turbulence is to treat the problem in the true quantum many-body context; study at first the dynamics of simple vortex configurations and extrapolate then to systems containing more vortices. This approach constitutes a bottom-up approach to the problem, putting the quantum dynamics at the heart of the picture. This paper treats the problem of understanding vortex dynamics with this second approach.

\begin{figure}
\centering
\includegraphics[scale=0.3]{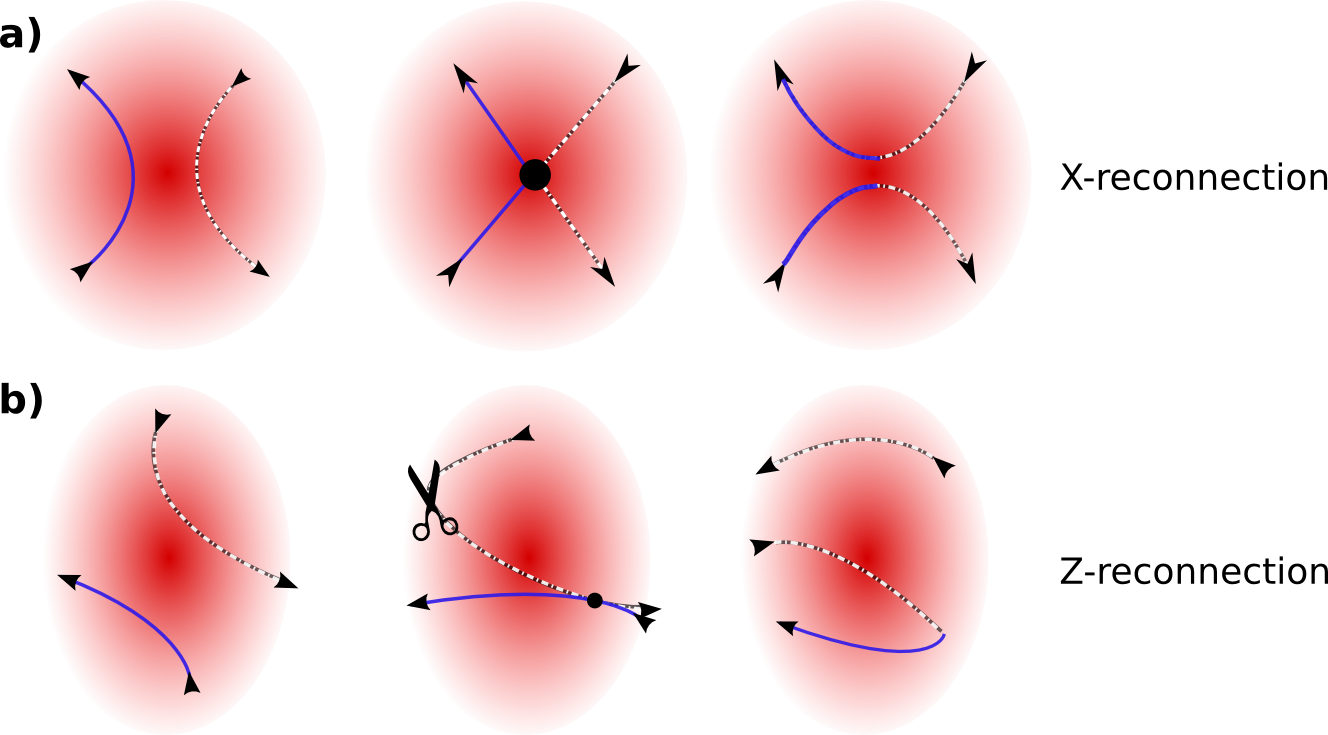}
\caption{Schematic depiction of two distinct topologies of the reconnection of the quantum vortices. Row a) shows the X-reconnection of the two vortices in a symmetric trap (the vortex lines reconnect in an X-shaped topology). It always happens in regions of higher density, i.e., at the center of the trap. Row b) shows the Z-reconnection in a gas that is asymmetrically confined ($\omega_y/\omega_x\geq 3/2$ and the vortex lines follows a Z-shaped topology). The reconnection in the latter case happens towards the edge of the gas.}
\label{sketch}
\end{figure}

The vast majority of previous studies of the three-dimensional vortex dynamics has been carried out in uniform systems. Experiments with trapped gases however necessitate simulations of vortex interactions in non-homogeneous systems. In the present work, we solve the many-body time-dependent Schr\"odinger equation (TDSE) as an initial value problem in three dimensions, for a state of the trapped gas containing two perpendicular vortices. This initial state can be created by various vortex imprint techniques (see for instance Ref.\cite{Zamora2012}). The many-body method used to solve the TDSE is the \textit{Multi-Configurational Time-Dependent Hartree for Bosons} (MCTDHB)\cite{MCTDHB1} and its recursive software implementation (R-MCTDHB)\cite{rmctdhb_web}. This method allows us to study the dynamical fragmentation of the system and look for beyond mean-field phenomena. 
%
%
We study the dynamics of the vortices for different values of the interaction strength as well as the trap anisotropy. Both parameters are of experimental relevance and expected to impact the dynamical evolution of the system and, more specifically, the time to the QVR. It is found that the QVR can follow two separate paths, depending on the anisotropy of the confinement: either through tail-exchange between the two vortices in the region of high density (isotropic or close to isotropic trap) or at the border of the cloud, in regions of low density (strongly anisotropic). The former is the standard way to QVR (see for instance Ref.\cite{Vinen2002}) and we call it herein `X-reconnection'. The latter is rather unseen before and termed `Z-reconnection' due to the peculiar shape of the  topology of the QVRs. A sketch of these distinct reconnection paths is shown in Fig.~\ref{sketch}.

The structure of the paper is the following. Section~\ref{Theory} introduces the system Hamiltonian and defines the quantities of interest that will be used. It outlines the vortex imprinting method that defines the initial state of the system. Section~\ref{Results} presents the results of the calculations and in Sec.~\ref{Conclusions} conclusions are drawn.

\section{System Hamiltonian and initial state} \label{Theory}

We study a system of $N$ dilute bosons at zero absolute temperature $T=0$ that are trapped in a harmonic potential in three spatial dimensions. We solve the time-dependent many-body Schr\"{o}dinger equation as an initial value problem where the initial state is fixed to posses two perpendicular vortex lines and evolve this state by propagating in real-time.
The TDSE reads
\begin{equation} \label{schrod}
i\frac{\partial}{\partial t} \Psi=\hat{H} \Psi,
\end{equation}
where the Hamiltonian is
\begin{equation} \label{eqmanybodyhamiltonian}
\hat{H} (\vec{r}_1,\vec{r}_2,...,\vec{r}_N)=\sum_{i=1}^{N} \hat{h}(\vec{r}_i) + \lambda_0\sum^N_{i<j}\hat{W}(\vec{r}_i-\vec{r}_j).
\end{equation}
Here $\vec{r}_i$ is the position of the $i$\textsuperscript{th} boson, $\lambda_0$ is the interaction strength and $\hat{W}$ is the interaction potential. The one-body Hamiltonian operator is $\hat{h}(\vec{r}_i) = -\frac{1}{2}\hat\nabla^2_{\vec{r}_i} + \hat{V}(\vec{r}_i)$ with
\begin{equation} \label{trap}
\hat{V}(\vec{r}_i)=\frac{1}{2} \omega_x {x_i}^2+\frac{1}{2} \omega_y {y_i}^2+\frac{1}{2} \omega_z {z_i}^2
\end{equation}
the operator corresponding to the confining trap. Our trap is characterized by the anisotropy on the $x-y$ plane which is quantified by the axis ratio (or anisotropy parameter): 
$$ 
\epsilon=\frac{\omega_y}{\omega_x},
$$ 
while $\omega_z$ is kept fixed at unity.
The interparticle interaction potential $\hat{W}$ is a Dirac-delta, representing a contact interaction.
Equation~\ref{eqmanybodyhamiltonian} is dimensionless and can be made dimensional by multiplying by the unit of energy $\frac{\hbar^2}{mL^2}$.\footnote{$L$ is chosen to be $L=1\mu{}m$, $m$ is taken to be the mass of an $^{87}Rb$ atom, $m=86.90918u$\cite{Rubidium}, such that the scaling factor corresponds to an angular frequency of $\omega=731{rad/s}$.}

The MCTDHB method assumes a general many-body ansatz for the solution of the TDSE which is an expansion over a set of many-body basis functions, also known as permanents\cite{MCTDHB1}. Each of the permanents describes a condensed or fragmented system and is time-dependent. In our case we choose $M=2$ orbitals and the expansion becomes:
\begin{equation}
\vert \Psi \rangle = \sum_{k=0}^{N} C_k(t) \vert N-k, k ; t\rangle.
\end{equation}
The permanents $|N-k,k;t\rangle$ are built over the two orbitals $\phi_1,\phi_2$ and, together with the coefficients $C_k(t)$, are determined variationally at each time $t$.
A guess wavefunction is relaxed into the ground state $\Psi_{\text{gnd}}$ by propagating in imaginary time until the energy is  converged. From $\Psi_{\text{gnd}}$ the orbitals $\phi_{1,\text{gnd}}, \phi_{2,\text{gnd}}$ are obtained. Their occupations are found to be close to $100\%$ and $0\%$ respectively. The initial state of two orthogonal vortices is constructed by defining a new wavefunction $\Psi$ using a modified $\phi_1$:
\begin{eqnarray}
\nonumber
\phi_1(\vec{r};t=0)= & \mathcal{N}~\phi_{1,\text{gnd}} \Bigg[
\tanh\left(\sqrt{\left(\frac{z-z_0}{
\sigma_z(z)}\right)^2+\left( \frac{y}{\sigma_y(y)}\right)^2 }\right) \exp\left(i
\tan^{-1}\left(\frac{y}{z-z_0}\right)\right) \\
           & + \tanh\left(\sqrt{\left(\frac{z+z_0}{
\sigma_z(z)}\right)^2+\left( \frac{x}{\sigma_x(x)}\right)^2 }\right) \exp\left(i
\tan^{-1}\left(\frac{x}{z+z_0}\right)\right) \Bigg],
 \label{iniState}
 \end{eqnarray}
The above state possesses one vortex parallel to the $x$-axis passing through $(0,0,+z_0)$ and a second vortex parallel to the $y$-axis passing through $(0,0,-z_0)$ at $t=0$. The parameters $\sigma_x(x) = \sigma_{x_0} e^{x^2/4}, \sigma_y(y) = \sigma_{y_0} e^{y^2/4}, \sigma_z(z) = \sigma_{z_0} e^{z^2/4}$ define the width of the vortex core. $\sigma_{x_0},\sigma_{y_0},\sigma_{z_0}$ are found, such that the energy obtains a minimum.
The values that we found and used for the different anisotropies are shown in Table \ref{table}. The exponential modulation follows from the fact that the vortex core at equilibrium is of the order of magnitude of the healing length $\xi$.
Since the latter is inversely proportional to the square root of the density of the gas, the vortex core is so too.
For the interaction strength chosen the cloud density profile is well approximated by a Gaussian shape so the vortices are expected to `open out' as we move towards the edge of the gas. The $\tanh$ functional form of the initial state of Eq.~\ref{iniState} was chosen such that the density vanishes along the lines $(z=z_0,y=0)$ and $(z=-z_0,x=0)$ while the complex-exponential terms create phase profiles that have quantized circulation about the vortex cores. It is important to note, that the vortex imprinting solely in the first orbital $\phi_1$ is sufficient in order to create the vortex pair in the many-body wavefunction:  since in the ground state of the parabolic potential $\Psi_{\text{gnd}}$ the first orbital occupation is larger than $99.9\%$ the wavefunction is almost fully determined by the orbital $\phi_1$ on which the vortices are imprinted.

Herein we choose $z_0=0.5$. The energy of the two-vortex state (Eq.~\ref{iniState}) as compared to the ground state with no vortices is found to increase by a factor of approximately $2$, for all choices of the interaction strength. Thus, energetically, our ansatz seems not to be a major disturbance to the ground state. 
The above initial state is propagated in real time and the quantities of interest are calculated.
In what follows, we define the density, occupation numbers, and reconnection time.

\begin{table}
 \begin{center}
  \begin{tabular}{| c || c | c | c | c | }
    \hline 
     $\epsilon$ & 1-1.4 & 1.5-1.9  &  2-2.9  &  3       \\ \hline
     $\sigma_{x_0},\sigma_{y_0},\sigma_{z_0}$ & 10,10,10 & 10,12,10 & 10,15,10 & 10,16,10  \\ 
    \hline
  \end{tabular}
 \end{center}
\caption{Values of the optimized vortex core parameters $\sigma_{x_0},\sigma_{y_0},\sigma_{z_0}$ (i.e. optimal values where the energy obtains a minimum) for $g=10$ and different anisotropies. For the isotropic cases $\epsilon=1.0$ and $g=0.1,1,100$ the values $\sigma_{x_0}=\sigma_{y_0}=\sigma_{z_0}=10$ were used. Precisely, for $\epsilon=1$, the energy obtains a minimum at infinitely large $\sigma_{x_0},\sigma_{y_0},\sigma_{z_0}$. However we imply this cut-of at $\{10,10,10\}$; beyond these values the energy drops marginally and the shape of the imprinted vortex does not change significantly.}
\label{table}
\end{table}

The one-body reduced density matrix (RDM) is defined as
\begin{eqnarray} \label{oneBodyRDM}
\rho^{(1)} (\vec{r} \vert \vec{r'};t)=N\int \vec{dr}_2...\vec{dr}_N
&\Psi^* (\vec{r'},\vec{r}_2,...,\vec{r}_N;t) &\Psi(\vec{r},\vec{r}_2,...,\vec{r}_N;t),
\end{eqnarray}
which can be decomposed into its eigenfunctions
\begin{equation} \label{orbitals}
\rho^{(1)} (\vec{r} \vert \vec{r'};t)=N\sum_{i=0}^M \rho_i^{(NO)}(t)
\phi_i^*(\vec{r}';t)\phi_i(\vec{r};t),
\end{equation}
where $\rho_i^{(NO)}(t)$ are the natural occupations and $\phi_i$ the natural orbitals. A many-body system is said to be condensed when $\rho^{(NO)}_k\thicksim N$ for some $k$, i.e. when one only natural orbital is macroscopically occupied\cite{Onsager1956}. When a finite number of occupation numbers $\rho_i^{(NO)}$ are of the order of $N$ then the system is called  fragmented\cite{Nozieres1982}. The \emph{density} in real-space is defined as the diagonal of the RDM $\rho(\vec{r},t)=$\\$\rho^{(1)}(\vec{r} \vert \vec{r'}=\vec{r};t)$. 
The numerically calculated density is a four-dimensional array and in order to visualize it we plot isosurfaces of constant $\rho$ at some given time. The numerical analysis is done with a discrete variable representation (DVR) on a grid of  $128^3$ grid-points, extending from $-8$ to $+8$ in each spatial dimension.

\section{Results} \label{Results}

\begin{figure}
\centering
\includegraphics[width=1\textwidth]{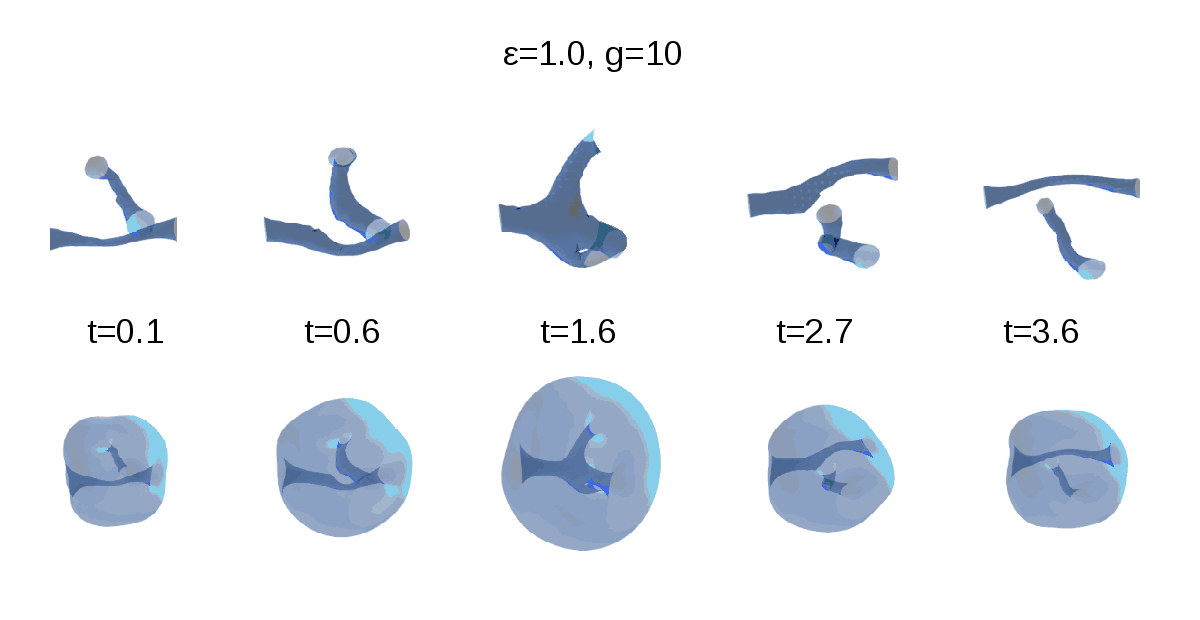}
\caption{Quantum vortex reconnections for $\epsilon=1$ (X-reconnection). Plots of the real-time dynamics for the symmetric trap, at different times. The top row of the figures is zoomed in to the bulk density and the bottom row shows the whole grid, plotted with an opacity of $0.25$. The isosurface values for each time point are defined by the minimum value of the density plus $0.05$ times the peak-to-peak value of the density. The vortices reconnect at the center of the trap where the density is maximal. The reconnection induces remarkable bulk density oscillations. The occupation numbers of the orbitals however remain almost unchanged. Video simulations of the dynamics for longer times are available\cite{youtube}.} 
\label{fig:reconnection}
\end{figure}

As descriptive quantities for the dynamics of the QVR we use the \emph{reconnection begin time} and \emph{reconnection end time}. These are defined as the first moment at which the isocontours connect and disconnect respectively. An error bar with magnitude of the time resolution (between $0.10$ and $0.25$) is ascribed to each data point, indicating that the densities were calculated at this time step. For example, in the simulation shown in Fig.~\ref{fig:reconnection} we find a vortex reconnection begin time of $t=0.70 \pm 0.10$ and a vortex reconnection end time of $t=2.70 \pm 0.10$.

\paragraph{Variation of the interaction strength.}
Dimensional analysis of the GPE reveals that one can define a ``healing time'' $\tau$, in direct analogy to the healing length $\xi$ of the condensate as $$\tau=\frac{m\xi^2}{\hbar}=\frac{m}{\hbar g \rho_0},$$ where $\rho_0$ is a representative constant value of the unperturbed density\cite{Yukalov2014}. 
In words, this is the time that it takes for the condensate to heal a perturbation over $\xi$. Hence, for increasing interparticle interaction $g$ the healing time of a perturbation decreases. This simple analysis suggests a scaling in time with $g$; the dynamics in systems that interact more strongly happen faster. This motivates the investigation of the role of the interaction strength, $g$, in the vortex dynamics and also of how the time-evolution of fragmentation of the system in time will be affected. 

In Fig.~\ref{fig:reconnection} an example reconnection can be seen. We plot the densities, as found from the solution of TDSE in real time, for the case $\epsilon=1,g=10$ at different times: before, during and after the reconnection. A standard X-reconnection is seen: the two vortices approach each other and they reconnect close to the center of the trap by exchanging tails.

Keeping the anisotropy fixed at $\epsilon=1$ (spherically symmetric trap) we now study the QVR for the values $g\in\{0.1, 1, 5, 10, 20, 50, 100\}$, where $g=\lambda_0(N-1)$ is the interaction parameter. Repeating the study of the QVR for all these values of $g$, we calculate the reconnection begin and end times for each case. Interestingly, we saw that the interaction strength plays no significant role in the vortex reconnection begin/end time for our choice of initial state (see Table~\ref{gScan}). Even though the total spatial expansion of the density is, as expected, larger for larger $g$ the differences in the profile of the vortices as they evolve in time is not noticeable. The dynamics of the QVR are \emph{not} significantly affected by the changes in the strength of the interparticle interaction, at least for the values of $g$ examined here, that span $4$ orders of magnitude.

\begin{table}
\begin{center}
\begin{tabular}{| l || c | c |}
  \hline
  $g$ & Reconnection begin time & Reconnection end time  \\ \hline
  0.1 & 0.60 & 3.0 \\ 
  1 & 0.60 & 3.0 \\
  5 & 0.65 & 3.0 \\
  20 & 0.65 & 3.0 \\
  50 & 0.60 & 3.0 \\
  100 & 0.60 & 3.0 \\
  \hline  
\end{tabular}
\end{center}
  \caption{Vortex reconnection times for a scan over $g$ values. These results show that the dynamics of the first reconnection are marginally changed for a range of $g$ values, from weak to moderately strong, over four orders of magnitude. The error on each time data point is $\pm 0.05$. 
  \label{gScan}}
\end{table}

It was found that, on the timescale of the first reconnection, the system does not fragment even for the larger interaction strength values (maximum $g=100$). This signifies that for the system parameters and initial conditions studied, the time dependent GPE would be a sufficient model for the time scales presented here. One can thus see that the QVR has no immediate effect on the orbital occupations.
For larger $g$ values ($g=10$ and $g=100$), we saw that system fragments only slowly over time. Indeed, in the case of $\epsilon=1.0$ and $g=10$ we have found that for times as long as $t=100$ the first natural occupation stays above $90\%$. In other words, for the time scale when the first approximately $30$ reconnections happen the gas stays close to a condensed state. Though, it could well be that for much longer times than that the state cannot be considered condensed anymore and quantum correlations become important and affect the QVR events.

\begin{figure}
\centering
\includegraphics[width=1\textwidth]{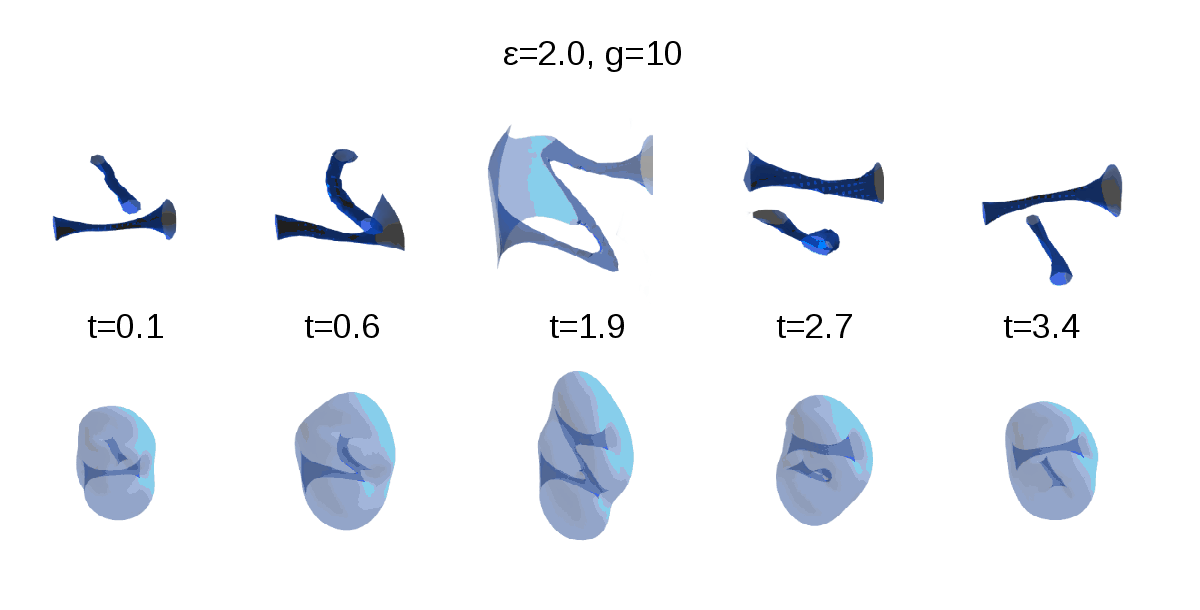}
\caption{Quantum vortex reconnections for $\epsilon=2$ (Z-reconnection). Plots of the real-time dynamics for the asymmetric trap, at different times. The vortices reconnect towards the edge of the trap where the density is minimal. As in the previous case, remarkable bulk density oscillations are seen in phase with the reconnections. The occupation numbers of the orbitals however remain almost unchanged and close to $100\%$ and $0\%$ for the first and second natural orbitals respectively, throughout the reconnection. Video simulations of the dynamics for longer times are available\cite{youtube}.} \label{fig:altreconnection}
\end{figure}

\paragraph{Variation of anisotropy.}
We now turn to the study of the dependence of the QVR on the trap anisotropy $\epsilon=\frac{\omega_y}{\omega_x}$.

For values $1.0\leq\epsilon\leq{}1.4$ the QVR happens through a similar mechanism to that of Fig.~\ref{fig:reconnection}, i.e. it shows an X-reconnection. However, for $\epsilon > 1.4$ it seems that the QVR is shifted towards the edge of the condensate. The vortices, due to the anisotropy, tend to align and they follow a different path: they reconnect at regions of low density, i.e. the edge of the cloud instead of the center of the trap. The whole process has a different topology (Z-reconnection) than the previously discussed case (X-reconnection). See also Fig.~\ref{sketch}, that highlights the difference.
Figure~\ref{fig:altreconnection} shows the Z-reconnection of the strongly anisotropic gas. One sees in the central image of this figure a `Z' shape, which occurs as one of the initial vortices has bent so much that the curved section has met the edge of the condensate bulk. This causes a break in the vortex line and forms one vertex of the `Z' shape. Simultaneously one end of the bent vortex meets the end of the other initial vortex to form the second vertex of the Z shape.

\begin{figure}[h!]
\centering
\includegraphics[width=0.8\textwidth]{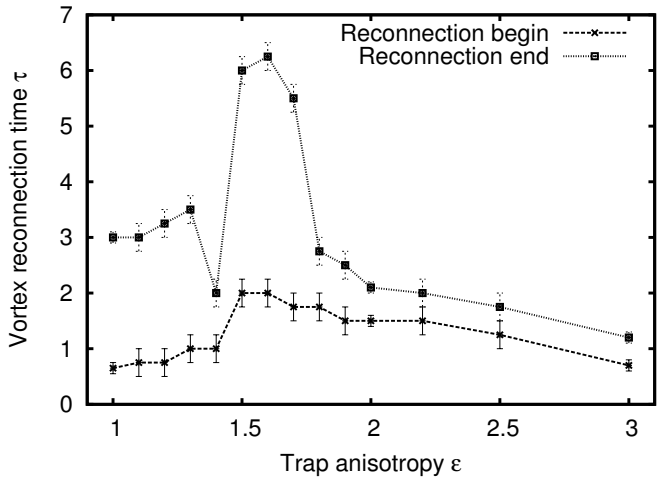} 
\caption{Quantum vortex reconnection times for varying axis ratio (i.e. trap anisotropy parameter) $\epsilon$. For values up to $\epsilon=1.4$ no significant change in the begin/end time are seen as $\epsilon$ increases. The jump in reconnection time from $\epsilon=1.4$ to $\epsilon=1.5$ is indicative of the change in reconnection type, from the X-reconnection shown in Fig.~\ref{sketch} a) to the Z-reconnection shown in~\ref{sketch} b) in which the QVR happens towards the edge of the cloud. For larger values of $\epsilon$ the vortices stay connected for shorter times.} \label{fig:escan}
\end{figure}

We repeated the calculations for various different values of the anisotropy parameter $\epsilon$. In Fig.~\ref{fig:escan} the first QVR begin and end times are plotted as a function of $\epsilon$. The change in the path of QVR, from X- to Z-reconnection creates a large change in the times associated with the reconnection, which can be seen between $\epsilon=1.4$ and $\epsilon=1.5$ in Fig.~\ref{fig:escan}. For large trap anisotropies the Z-reconnection path remains as the reconnection method and the reconnection events happen faster as the edge of the cloud is pushed closer to the position of the initial vortex line.

A common feature that all the simulations share is the bulk oscillations during the QVR: the cloud expands and reaches a maximum at the moment of the QVR, seemingly \emph{in-phase} with the QVR.  Furthermore, calculations for times as long as $t=140$ suggest that the repeating reconnections persist and occur indefinitely. In Ref.\cite{youtube} the reader can see the videos of the real-time dynamics for the cases of $\epsilon=1.0, 2.0$ and $3.0$.

\section{Conclusions} \label{Conclusions}

From the results presented in this work, three main conclusions can be drawn. The first is that there is no direct connection between vortex reconnection and system fragmentation for the coherent initial state containing two orthogonal vortices. We found no significant deviations in the orbital occupations at the short time scales at which the vortex reconnection occurs. For long times the occupation of the first orbital settles at approximately little less than $90\%$. The second conclusion is that reconnection begin and end times are unaffected by variations of the interaction strength. Hence the QVR dynamics do not appear to scale in time with the interaction. In other words, as the interaction increases and the density at the center decreases the velocities of the vortices do not change. One could ask if the same behaviour persists for larger $g$ and very strongly repulsive gases. Independently of what this answer be, we believe that this counter-intuitive result deserves more attention and is the subject of future work.
The last principal conclusion is that different anisotropies can lead to different routes to reconnection which significantly change the system dynamics. This is shown in Figs.~\ref{fig:reconnection} and ~\ref{fig:altreconnection} and in the videos in Ref.\cite{youtube}. It was also found that the vortex connections will occur repeatedly for long times: approximately $40$ QVR happen during the a time that is equal to $20$ trap periods.

Among a handful of publications that deal with vortex reconnections in trapped BECs, the present paper is the first research into the many-body nature of vortex dynamics and reconnections in three-dimensional trapped ultracold Bose gases. Furthermore it is, to the best of our knowledge, the first work to numerically study vortices in highly asymmetric traps, as well as being the first work to study quantum vortices in a many-body context in three spatial dimensions. As a next step, it would be appealing to study the dynamics of \emph{initially fragmented} vortices and look for many-body mechanisms of vortex reconnection, if any, that go beyond the commonly applied Gross-Pitaevskii mean-field method. We hope that the present work will stimulate experimental research.

\begin{acknowledgements}
The authors wish to thank FAPESP for financial support. A.U.J.L. acknowledges financial support by the Swiss SNF and the NCCR Quantum Science and Technology. T.W. thanks DAAD for financial support. Computational time in the Hermit Cray computer of the High Performance Computing Center in Stuttgart is also gratefully acknowledged.
\end{acknowledgements}

\pagebreak

\end{document}